# X-ray Timing of Neutron Stars, Astrophysical Probes of Extreme Physics

Z. Arzoumanian, S. Bogdanov, J. Cordes, K. Gendreau, D. Lai, J. Lattimer,
B. Link, A. Lommen, C. Miller, P. Ray, R. Rutledge, T. Strohmayer,
C. Wilson-Hodge, and K. Wood

February 15, 2009

**Contact information:**

Zaven Arzoumanian, CRESST-USRA and X-ray Astrophysics Laboratory
Code 662, NASA Goddard Space Flight Center, Greenbelt, MD 20771
301-286-2547 / Zaven.Arzoumanian@nasa.gov

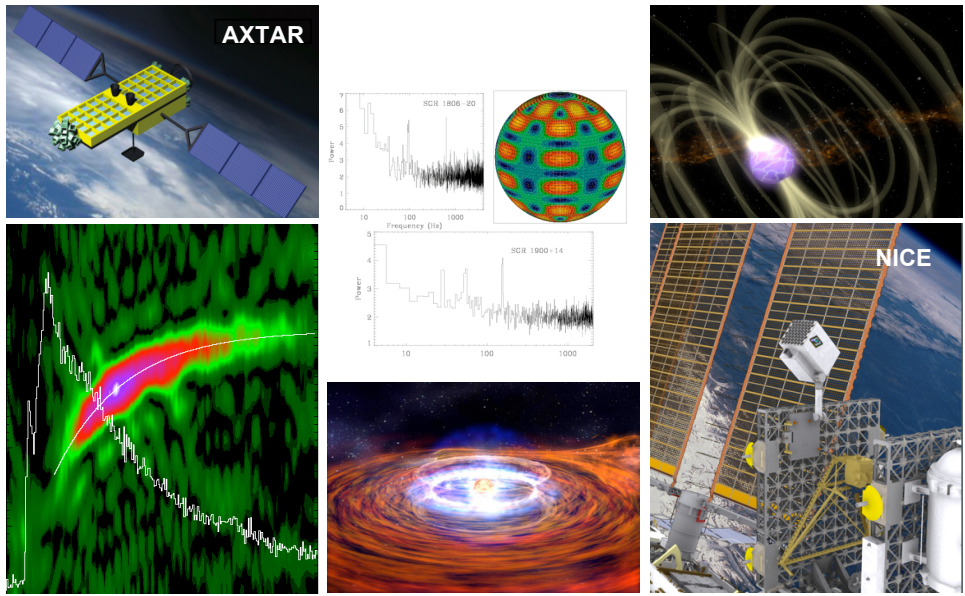

**Related Astro2010 white papers:**

J. Cordes et al., *Tests of Gravity and Neutron Star Properties from Precision Pulsar Timing and Interferometry*

P. Freire et al., *Constraining the Bulk Properties of Dense Matter by Measuring Millisecond Pulsar Masses*

D. Lai et al., *Extreme Astrophysics with Neutron Stars*

F. Paerels et al., *The Behavior of Matter Under Extreme Conditions*

J. Tomsick et al., *X-ray Timing of Stellar Mass Black Holes*

# Introduction

The characteristic physical timescales near stellar-mass compact objects are measured in milliseconds. These timescales—the free-fall time, the fastest stable orbital period, and stellar spin periods—encode the fundamental physical properties of compact objects: mass, radius, and angular momentum. The characteristic temperature of matter in the vicinity of a neutron star or black hole is such that the principal electromagnetic window into their realms is the X-ray band. While it was understood for decades that the measurement of such timescales would provide unique insights into compact stars and their extreme physical environments, it is only in the past decade that modulations in X-ray flux on millisecond timescales have been detected from accreting neutron stars and black holes. NASA's *Rossi* X-ray Timing Explorer enabled the discoveries of millisecond pulsations that reveal the spin rates of neutron stars, and quasi-periodic oscillations (QPOs) at frequencies consistent with the orbital motion of matter in the strongly curved space-time of both neutron stars and black holes (see [1, 2] for reviews). For non-accreting systems, the rapid intensity variations of radio pulsar signals were historically the key to identifying their origin with neutron stars. With the advent of X-ray astronomy, the stars themselves were detected through their thermal emissions. Because of these connections to the fundamental properties of neutron stars, X-ray timing studies remain today the most direct means of probing their structure and dynamics. While current X-ray observatories have revealed many relevant and fascinating phenomena, they lack the sensitivity to fully exploit them to uncover the fundamental properties of compact objects and their extreme physics.

With this white paper, we summarize and highlight the science opportunities that will accompany an order-of-magnitude improvement in X-ray timing sensitivity, a goal attainable in the coming decade. We focus primarily on neutron stars, as a companion white paper (Tomsick et al.) will emphasize black hole science. Our scientific objectives can be formulated broadly as a pair of questions: *i*) What is the nature of matter under conditions of extreme density and magnetic fields? *ii*) What is the end state and mass of a star as a function of its initial mass, composition, rotation, and any stellar companions?

# Neutron stars: matter at the limit of finite density

Neutron stars embody extremes inaccessible anywhere else in the Universe, but two insights provide the fundamental physics context:

- According to current understanding, neutron stars represent the strongest gravitational environment in which matter of any kind can stably exist. An incremental addition of matter would drive a massive neutron star beyond the point at which it could support its own weight, and the star would collapse to a black hole.
- The state of cold, stable matter at ultra-high density remains one of the most important unsolved problems in subatomic physics. Neutron stars represent a density-temperature regime that can be explored in no other way.

A neutron star's interior structure is captured, in a global sense, by the still-uncertain equation of state (EOS) of bulk nuclear matter. The EOS relates density to pressure within the star (Figure 1) or, through General Relativity, its mass $M$ to radius $R$. Most EOSs predict that $R$ will shrink as $M$ grows and the self-gravitational force increases, but different assumptions about interior composition produce different detailed mass-radius relations. Thus, measurements of $M$ and $R$ probe dense matter. The conditions resulting from the enormous pressure at the center of the star may include *i)* dissolution of individual neutrons into an undifferentiated soup of quarks and gluons;

*ii)* a phase transition to a "Bose condensate" of pions or kaons; or *iii)* a phase transition to yet-more-exotic matter made up of hyperons. Which, if any, of these theoretically well-motivated outcomes actually occurs in Nature is unknown—the interior composition of neutron stars has remained elusive since the first theoretical description of these objects by Oppenheimer & Volkoff (1939; [3]), through their discovery by Hewish et al. (1968; [4]), and to the present day.

The surface layers of neutron stars present unique physics of their own: while the EOS is a static property, phenomena associated with the accreted ocean and crust are primarily responsible for the dynamic nature of neutron stars, including starquakes, thermonuclear explosions, and interactions with the core that may involve superconductivity and superfluidity on stellar-mass scales.

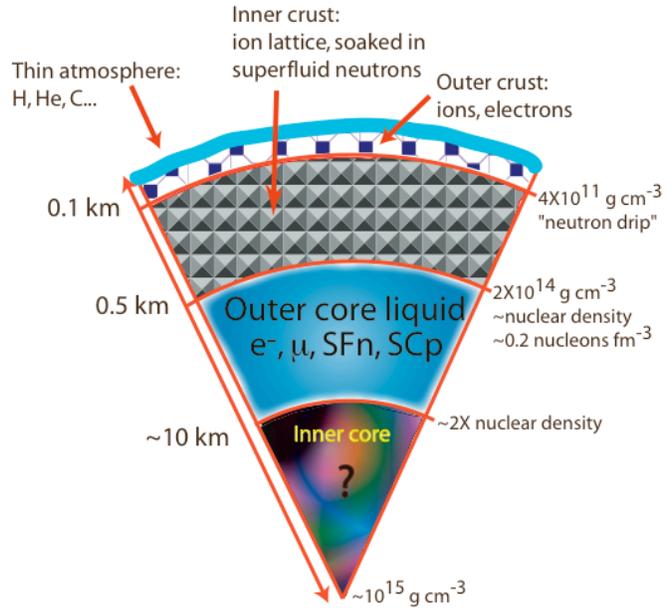

**Figure 1**. Current theoretical understanding of the interior composition of neutron stars. Uncertainty increases with depth.

A distillation of the fundamental questions of neutron star science, the measurements required to address them, and some of their broader implications, appear in Table 1. These objectives can be met in the coming decade through coordinated multi-wavelength efforts, with radio, X-ray, and gamma-ray observations providing, with the support of theory, complementary pieces of the puzzle (see, e.g., white papers by Cordes et al. and Freire et al. on mass measurements and other EOS constraints available through radio pulsar timing). The scientific environment is especially promising for studies in 0.2–20 keV X-rays, the electromagnetic band into which neutron stars, and the relativistic plasma processes around them, radiate significant fractions of their thermal, magnetic, and rotational energy stores. (For additional discussion of neutron star astrophysics, see white papers by Lai et al. and Paerels et al.)

**Table 1.** Fundamental questions of neutron star structure and dynamics.

| Science Questions | Measurements Needed | Implications |
| --- | --- | --- |
| What is the nature of ultra-dense matter in the interiors of neutron stars? | The mass-to-radius ratios of several neutron stars to ±5%. | Discriminate among proposed EOSs; constrain a basic unknown of nuclear physics, the nuclear symmetry energy. |
| What is the physics responsible for the dynamic behavior of neutron stars? | Characterization of outbursts, oscillations, and rotational irregularities. | Constrain the bulk properties of dense matter. Probe quantum phenomena in neutron stars. |

## Structure: the Equation of State of Neutron Stars

Above half the standard nuclear saturation density (the average density inside atomic nuclei, $\rho_s \sim 2.5\times10^{14}$ g cm$^{-3}$), nuclei give way to uniform nucleonic matter. Above about $2\rho_s$, many models of dense matter predict that exotic phases such as hyperons, kaon- or pion-condensates, or deconfined quark matter appear. Whatever the state of this matter, it exists in abundance in the millions of neutron stars that inhabit the Galaxy. The presence or absence of exotic phases within their cores (the "?" in Figure 1) has profound implications for their structure and evolution, and for their abundance relative to black holes. Discriminating among the many alternatives probes both the astrophysics of stellar endpoints and the basic physics of nuclear matter.

## Masses and Radii

Two of the outstanding issues surrounding neutron stars are also the simplest to state: What is the maximum mass of a neutron star? And what is a typical radius? These essential properties reflect different aspects of the dense matter EOS. A consequence of General Relativity, the maximum mass is controlled by the composition and stiffness of the EOS well above $\rho_s$. On the other hand, the neutron star radius is most influenced by the EOS in the vicinity of 1–2 times $\rho_s$; it is a direct probe of the unknown density dependence of the nuclear symmetry energy, the energy difference between pure neutron matter and symmetric (having equal neutron and proton abundances) matter [5]. Therefore, determination of both important parameters would set powerful constraints on the entire range of the dense matter EOS [6].

A few dozen neutron stars have had masses reliably determined (some to ±0.001 $M_\odot$) from binary pulsar timing. All measured masses are consistent at 4σ with a maximum of 1.5–1.65 $M_\odot$. In contrast, no reliable radius measurement is currently available, although most estimates are consistent with the expected theoretical range of 10–15 km. Of particular importance is the fact that no neutron star has had both its mass and radius determined. Nuclear theory predicts distinct mass-radius (*M-R*) relationships for given models of particle composition and interaction in dense matter [5]. These predictions must be confronted with measurements, to isolate models that are consistent with observed neutron stars and to rule out large numbers of alternative EOSs. Represented graphically (Figure 2), mass and radius constraints for a number of neutron stars will define bounded regions in the *M-R* plane. EOS curves that do not pass through *every* region must be ruled out; any that do will remain viable (e.g., [7]). Precise masses may be obtained from pulse timing—in the radio or X-rays for neutron stars in binary systems—but direct radius measurements are extremely challenging for objects just kilometers across at astronomical distances. A different approach is needed, one that simultaneously constrains both *M* and *R*. Such an approach is available thanks to decades of theoretical work to define the mass and radius impli-cations of many observable phenomena (Table 2; see, e.g., [8]): a pair of precise measurements (of redshift and $R_\infty$, for example) from a single target solves for the two unknowns, $M$ and $R$. If the mass is already known from timing studies, just one additional measurement is needed. By characterizing these phenomena with unprecedented precision, distinct "allowed" regions in $M$-$R$ for a number of neutron stars can be produced with which to constrain the EOS.

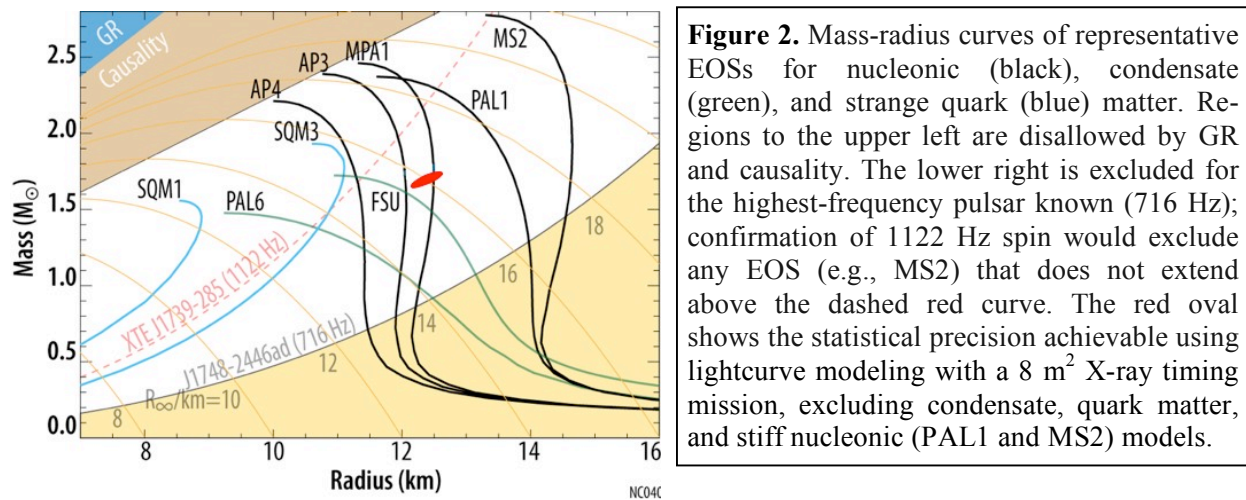

**Figure 2.** Mass-radius curves of representative EOSs for nucleonic (black), condensate (green), and strange quark (blue) matter. Regions to the upper left are disallowed by GR and causality. The lower right is excluded for the highest-frequency pulsar known (716 Hz); confirmation of 1122 Hz spin would exclude any EOS (e.g., MS2) that does not extend above the dashed red curve. The red oval shows the statistical precision achievable using lightcurve modeling with a 8 m² X-ray timing mission, excluding condensate, quark matter, and stiff nucleonic (PAL1 and MS2) models.

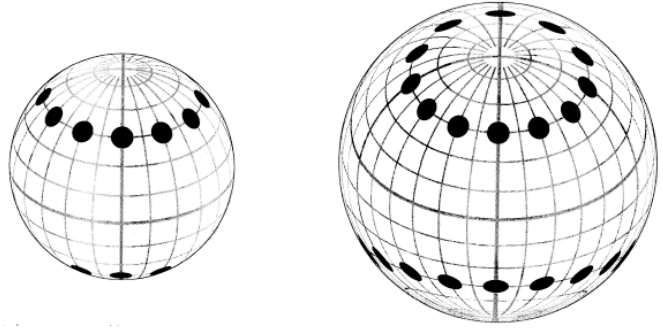

**Figure 3.** A rotating neutron star with a pair of antipodal hot-spots as seen by a distant observer in the absence of light-bending (left) and for strong gravity, $\beta = 0.25$ (right). From Nollert et al. (1989).

**Table 2.** Expressions linking mass and radius to measurable phenomena.

| Measurement | M, R dependence | Approach |
|---|---|---|
| Redshift/compactness | $\beta = GM/Rc^2$ | Lightcurves and spectra |
| Surface gravity | $g = GM/R^2$ | Lightcurves and spectra |
| Light-bending magnified radius | $R_\infty = R/\sqrt{1 - 2GM/Rc^2}$ | Thermal spectra |
| Inner edge of accretion disk | $R_{disk} \geq R$ | Broadened Fe lines |
| kHz QPO frequency (one of several theoretical relations) | $\nu_{QPO} = \sqrt{GM/4\pi^2 R_{disk}^3}$ | Fast timing of X-ray binaries in outburst |
| Maximum mass | $M \leq M_{max}$, for all $R$ | Pulse timing |
| Minimum spin period | $P_{min} \propto \sqrt{R^3/M}$ | Pulsation searches |
| Fractional moment of inertia in crustal superfluid | $\Delta I/I \propto R^4/M^2$ | Glitch monitoring |
| Seismic vibrations | Mode-dependent | Flux oscillations in flares, bursts |

Some classes of neutron stars are faint, with the total X-ray flux the sum of several emission components with distinct spectral and temporal properties. Without sensitive telescopes, measurements are hampered by the need to increase signal-to-noise ratios by either *i)* averaging photon energy spectra over time, or *ii)* averaging temporal lightcurves over energy, when in fact emission processes are best investigated as a function of energy and time simultaneously.

**Lightcurve modeling** has been demonstrated for thermal X-ray pulsations from nearby millisecond pulsars (MSPs), persistent accretion-powered pulsations from neutron stars in binary systems, and X-ray burst oscillations. *A key goal in the next decade is to enable such modeling to precisely measure masses and radii*, to achieve the standard set by nuclear theory for distinguishing between proposed nucleonic and exotic core models: ±5% uncertainty on radius measurements. Relativistic effects set by $M$ and $R$ strongly influence the propagation of photons from the surfaces of neutron stars, especially those that rotate rapidly. Gravitational light bending, which depends on $GM/Rc^2$ (e.g., [9]), brings the geometrically obscured "back side" of a star into view (Figure 3), magnifying its apparent radius and altering the amplitude of pulsations for emission from rotating "hot-spots" on the surface. Similarly, the shape of pulses is affected by the rotational velocity of the emitting surface, which depends on $R$. Thus, detailed modeling of pulse profiles can be used to constrain masses and radii.

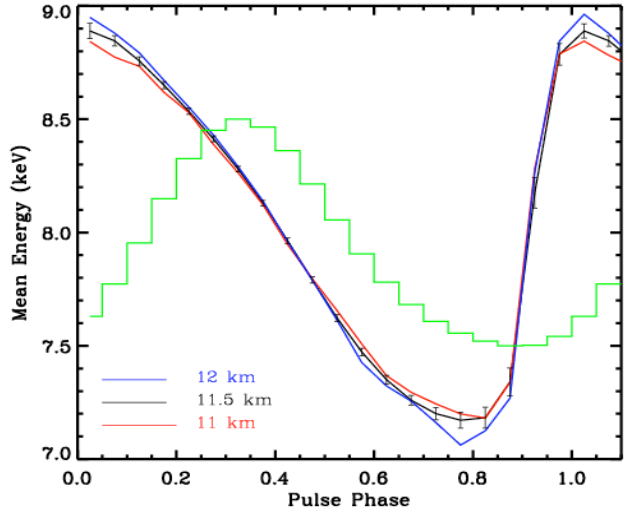

**Figure 4**: Simulated measurements from burst oscillations during the rising phase of X-ray bursts, appropriate for a mission concept similar to AXTAR. The red, black and blue curves show the variation with pulse phase of the mean of the energy spectrum due to rotational Doppler shifts, while the green histogram shows the pulse profile. Error bars on the black curve show the characteristic statistical precision achievable with 5 typical X-ray bursts from the LMXB 4U 1636–53.

Millisecond pulsars (MSPs) are ideal for this approach ([10], [11]): their atmospheres [12] are unmagnetized and thus simply modeled (simpler even than atmospheres of main-sequence stars), and they appear frequently in binary systems, providing independent mass measurements. Light-curve modeling is also promising for accretion-driven millisecond pulsars (AMSPs) and burst-oscillation sources [13, 14]. At the start of some bursts, high amplitude oscillations are observed that are produced by rotational modulation of the localized hot-spot created at the onset of nuclear burning. Both theoretical and observational clues indicate that the hot-spot can remain well localized for hundreds of rotation periods [15, 16]. This provides a relatively simple "system" to model and in which to measure the surface velocity of the star by detection of rotational-phase dependent Doppler shifts in the pulsed emission (Figure 4). For known spin frequency, the magnitude of the Doppler shift is proportional to the stellar radius; moreover, modeling of the pulse profile gives a measurement of the compactness, $M/R$. Thus, high signal to noise measurements of burst oscillations can, in principle, provide sufficient information to measure both $M$ and $R$ (Figures 2 and 4).

**Broad iron lines** have recently been discovered in four neutron-star low-mass X-ray binaries. The broadening is likely due to Doppler shifts at the innermost edge of the accretion disk, where orbital periods are measured in milliseconds and gravitational redshift is important. Spectral modeling yields the radius of the disk's inner edge, an upper limit on the star's radius. Moreover, because kHz quasi-periodic oscillations (QPOs) also probe the disk's inner edge (Table 2; [1]), the spectral and QPO radius estimates should agree, and track one another with changing accretion rate, yielding a mass estimate and testing both the QPO and disk spectrum models. Combined QPO and line-profile diagnostics establish regions in the $M$-$R$ plane [17] distinct from those of other phenomena.

**Absorption lines**, broad (~200 eV) features near 1 keV, are seen in six non-accreting neutron stars [18]. Probably atmospheric, it is not known whether the lines represent atomic transitions or cyclotron absorption. Once identified, their redshifts yield $M/R$ if they originate near the surface. Rotational-phase-resolved spectroscopy enabled by large collecting areas can reveal line origins.

**Pulsation searches** uncover the most consequential observable property of any neutron star: rotation rate. The fastest known spins suggest that these objects, for some EOSs, are on the verge of flying apart [19]. Searches with unprecedented sensitivity to rapid rotators could rule out an entire class of models that predict large, low mass neutron stars (Figure 2). Even if pulsation periods don't directly constrain the EOS, their discovery is important for eventual lightcurve modeling, characterization of orbits and masses, and for studying the origins of QPOs.

## Dynamics: the Crust and Quantum Fluids

Anticipated X-ray instrumentation will improve dramatically the observational characterizations of phenomena that shed light on the bulk properties and physics of neutron star crusts. The fundamental composition of crusts is not in doubt: as the density increases, nuclei become richer in neutrons, finally sharing volume with free neutrons above the "neutron drip" density of $\approx 4 \times 10^{11}$ g cm$^{-3}$. The density-dependent symmetry energy determines the thickness of the crust, the frequencies of seismic modes it can support, and neutrino cooling rates. The bulk properties of the crust, however, are not constrained well by existing observations even though they are implicated in many astrophysical phenomena. Examples of relevant questions are:

- What are the mechanical properties of the crust? What are its thermal properties?
- Do the crust and core interact through quantum liquids, superfluid neutrons and superconducting protons (e.g., Dean & Hjorth-Jensen 2003)?

### *Mechanical Properties*

**Asteroseismology.** Torsional vibrations of the neutron star crust are the likely origin of rapid X-ray flux modulations in giant flares from two soft gamma-ray repeaters. Analysis of such modes allows a measurement of the crust thickness, as well as constraints on magnetic stresses and the EOS [20, 21]. New X-ray telescopes will have the fast timing and low deadtime capabilities needed to detect oscillations up to tens of kHz with high throughput. In addition to giant flares, searches for seismic vibrations stimulated by lower-energy events, such as bursts and glitches in anomalous X-ray pulsars (AXPs) and rotation-powered pulsars, will be possible.

**Glitches**, sudden spin-ups of stars that are otherwise slowing down, are valuable probes of their interiors [22]. Glitches in AXPs suggest that, relative to radio pulsars, a large fraction of their moment of inertia is in superfluid, implying systematic structural differences [23].

### *Thermal Properties*

**Transient heating following glitches** is expected: current models suggest that glitches deposit ~$10^{42}$ ergs at the crust-core boundary (e.g., [24]). Because the thermal response time depends strongly on depth, the EOS, and the transport properties of the inner crust [25], spectral and lightcurve measurements on short and long timescales will provide unique insights. Failure to detect a thermal transient would, by itself, rule out a "quark star" nature for the glitching object.

### *Crust-Core Interactions*

**Timing noise and the ultimate clock stability of MSPs.** Long term timing of pulsars has shown that most, especially the youngest, exhibit clock instabilities of unknown origin [26]. Theories attempting to explain "timing noise" have generally looked toward internal causes, such as coupling between the core and crust. X-ray timing of MSPs, especially those with non-thermal pulsations, can provide sub-μs precisions competitive with radio timing, with the advantage that X-rays are immune to propagation delays in the ionized interstellar medium.

**Precession** of the spin axes of some isolated neutron stars must be due to non-spherical figures or internal dynamics. Non-sphericity has important ramifications for the material or magnetic stresses in the crust and for gravitational radiation. Crust-core interactions present a puzzle: observed precession timescales range from ~6 hours to several hundred days, but current understanding requires that any precession occur at a frequency orders of magnitude faster, seriously challenging current notions of the state of the outer core [27].

## Observational Requirements

The key requirements to realize these opportunities are large collecting area (~4–8 m$^2$), fast timing (≤100 μsec), and moderate spectral resolution ($E/\Delta E \sim 25$) in the ~0.2–20 keV X-ray band. The Advanced X-ray Timing Array (AXTAR), currently being studied as a MIDEX-class mission, would provide sufficient collecting area to achieve ~5% radius constraints based on the burst oscillation technique [28]. A low-cost SMEX Mission of Opportunity, the Neutron star Interior Composition Explorer (NICE) has also been proposed, providing an excellent match to the softer spectra of MSPs, to achieve ±5% radius measurements for 2–3 targets. The High Time Resolution Spectrometer detector planned for IXO will revolutionize the field by providing radius estimates for a large number of accreting and rotation-powered neutron stars.

Historically, effective areas for X-ray timing have climbed, and shorter timescales have been probed, with continual revelation of new phenomena. Discovery space in the form of the shortest accessible timescales accrues favorably, as area$^{-1}$, not area$^{-1/2}$. Increased area also means access to lower levels of modulation: fainter QPOs and flares, better definition of burst or eclipse profiles. A parallel trend is movement beyond simple periodic phenomena to complex variability patterns that now pose a challenge to the most advanced fluid dynamics. The world of X-ray timing is increasingly a place where gas, radiation, Fermi, and magnetic pressures act on fluids in strong gravity and conditions that promote nuclear reactions. Building larger collecting apertures will continue to bring dramatic benefits; if large areas are prohibitive for some reason, then greater *exposure* (area × time) can accomplish much, especially when combined with other capabilities. Excellent new technological candidates are begging for application to these challenges.